# Design Methodology for a Medium Voltage Single Stage LLC Resonant Solar PV Inverter

Parthkumar Bhuvela
Department of Electrical Engineering
University of South Carolina
Columbia, SC, 29208, USA
pbhuvela@email.sc.edu

Hooman Taghavi
Department of Mechanical Engineering
University of South Carolina
Columbia, SC, 29208, USA
htaghavi@email.sc.edu

Adel Nasiri
Department of Electrical Engineering
University of South Carolina
Columbia, SC, 29208, USA
nasiri@mailbox.sc.edu

*Abstract*—An inverter is generally employed with MV LFT to connect to the grid in a grid-tied PV system. However, in some single-stage topologies, the LFTs are replaced by HFT combined with an unfolder inverter. Generally, these topologies have limited use at high-power MV grids due to high switching losses on the primary side. This study proposes an LLC resonant converter-based single-stage inverter design procedure. Resonant converters make use of ZVS to reduce switching losses. The design includes both the resonant tank as well as output filter components. The design is verified by simulations in MATLAB/Simulink for various loads and input voltages at 13.8kV grid output voltage. THD simulations validate the filter design.

## I. INTRODUCTION

Renewable energy systems have attracted many researchers' attention in recent years. Effects on the environment from conventional energy resources and increasing energy demand have been significant factors that motivate the innovation of sustainable and diverse energy sources. This has led to studies to improve the small as well as large-scale solar photovoltaic (PV) inverters and their control designs [1]. Grid-tied solar PV inverter studies consist of single and multi-stage PV inverter topologies connected to a Low voltage grid. To connect to a medium voltage (MV) grid, Line Frequency Transformers (LFT) are utilized to match voltage levels and provide Galvanic isolation [1, 2]. This includes single stage as well as multi-stage converters. However, the large size, weight, high core losses, and cost of LFTs have driven recent studies involving topologies with two stages that typically include high-frequency transformer (HFT) embedded into the DC-DC converter stage that boosts the voltage levels, followed by an inverter stage HFT's smaller core size exhibit reduction of core losses and further loss reduction can be achieved by using topologies such as Dual Active Bridge (DAB) and Resonant Converters to reduce losses with soft switching techniques [3-7]. However, high switching losses can occur at the inverter stage due to medium to high frequency switching at MV. For these topologies to be used at high power medium voltage applications, the inverter stage must decrease the switching frequency to avoid high switching losses. This reduction in switching frequencies can increase the value of filter inductance and capacitance, which can increase the size of the overall system. Moreover, traditional single phase-shift (SPS) in DAB can lead to high circulating current in the HFT and increase losses at high loads. This leads to complicated optimizations in the control [4, 8-10]. Modular and double-stage inverters have a high number of switches and switching losses, which can lower the efficiency at high power, as suggested in [5, 11]. [12-14] suggest a single-stage inverter with flyback and buck-boost isolated topologies that connect to unfolding inverters. The unfolding inverter operates at line frequency; hence, the losses are significantly reduced at the unfolder stage. However, high voltage stress on switches and switching losses in the converter create major roadblocks for these topologies to be used at high power and MV. In [15], an Input Parallel Output Series (IPOS) LLC converter topology is studied to achieve MVDC. LLC converters make use of soft switching techniques to decrease switching losses. In [3, 16, 17]. Single-stage LLC converter-below voltages are proposed and verified with simulation and experiments at low-voltage. [18] uses a half-bridge LLC converter with a similar topology and produces 50kW on load resistance. However, these studies lack MV grid connection. Moreover, the design flow of the system adds complex optimizations. The system's output filter design is also essential in such topology, as the filter resonance can interfere with the resonance frequencies of the converter if not designed properly. This study proposes a simple design methodology for all the single-stage MV grid-tied inverter topology components. In Fig.1, the working of the converter is depicted. An IPOS system with two LLC resonant converters produces MV rectified AC current at the output capacitor bank of the converter. Each converter generates rectified sinusoidal voltage. This gets inverted using an Unfolder inverter operating at line frequency to neglect the switching losses at MV, making it suitable for grid tied MV applications. Studies on such systems with pseudo-DC link unfolder inverters have been limited to low-voltage (LV) and low-power applications so far. Appropriate design of resonant tank components is necessary for LLC resonant converters. With proper design, the leakage inductance of HFT can be used as a resonance inductor, eliminating the need for additional magnetics [19-21]. LLC converter gain function depends on the switching frequency and the amount of load. Hence, the proper choice of corner frequency for output filter design becomes another critical factor. The proposed system design methodology takes this into account as well. The MV system design and control are verified using MATLAB/Simulink simulations and compared with different design parameters. The THD of the system is simulated as well.

## II. LLC Converter Components

The LLC converter is typically controlled using switching frequency modulation or phase-shift modulation. To produce a rectified sinusoidal wave at the output, phase-shift modulation is used to get the output voltage gain all the way to zero. However, achieving Zero Voltage Switching (ZVS) is impossible over the entire output voltage range. Hence, frequency modulation is used as well to extend the ZVS region. As ZVS boundary for phase-shift modulation is given by equation (11), where θ is the input impedance angle, and φ is the phase-shift angle [19]. The impedance angle increases with an increase in the switching frequency, and the rate of change increases with an increase in power, as shown in Fig. 2b. Hence, the phase-shift angle ZVS boundary increases with an increase in switching frequency for a single LLC converter. In 2a, the gain vs. switching frequency for different loads is depicted. Here, a 1MW three-phase load pertains to a 1/3 MW single phase and 1/6 MW for a single converter load. The voltage gain for frequency modulated LLC Converter (M) and impedance angle (θ) using First Harmonic Approximation (FHA) are given by,

$$M = \frac{NF_x^2(m-1)}{\sqrt{(mF_x^2-1)^2 + F_x^2(F_x^2-1)^2(m-1)^2 Q^2}} \quad (1)$$

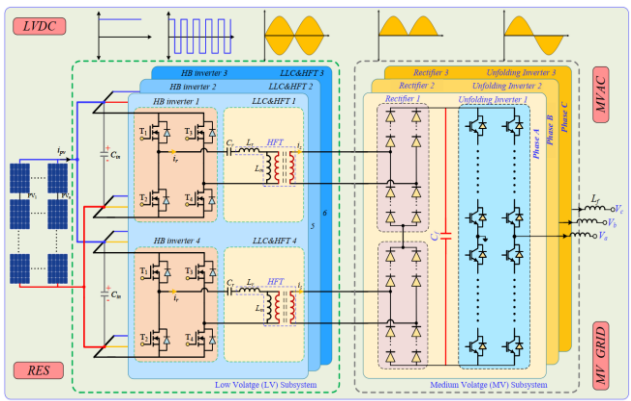

Fig. 1. LLC Converter-based Single stage PV inverter topology

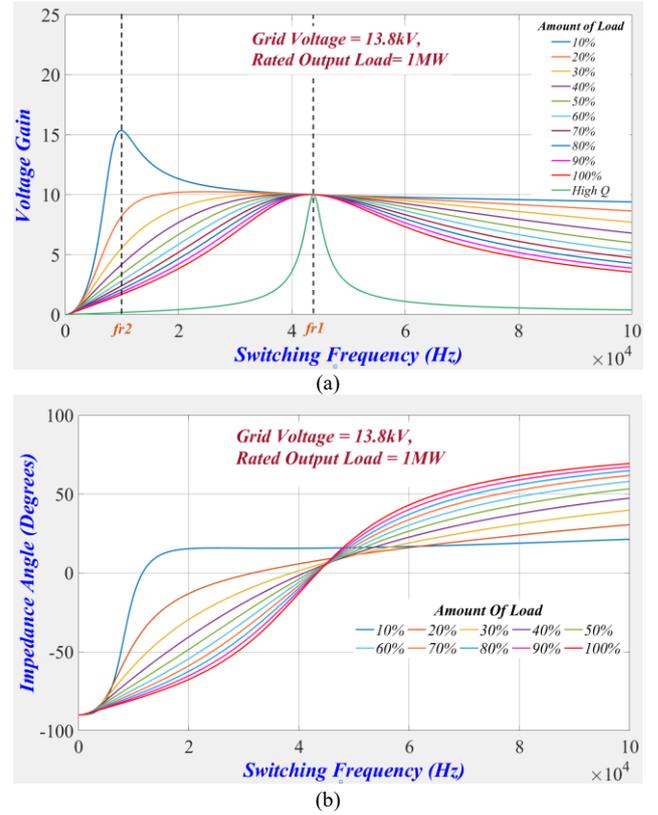

Fig. 2. (a) Voltage gain, and (b) Input Impedance angle vs. Switching frequency graph with Various load levels for a single LLC converter.

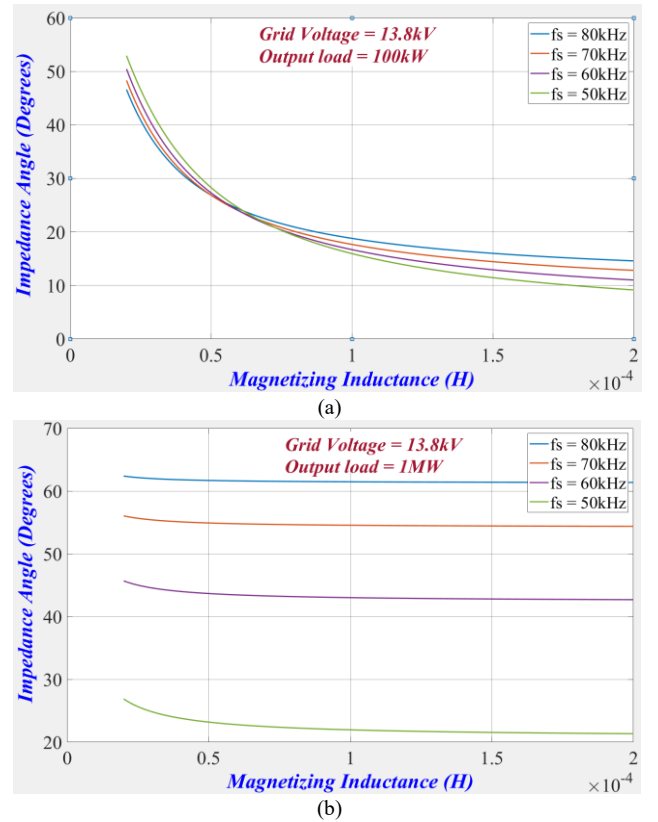

Fig. 3. Impedance Angle vs. Magnetizing Inductance for (a) low and (b) high load at different switching frequencies.

$$\theta = tan^{-1}(\frac{F_x^4 Q + F_x^2 L_x^2 + F_x^2 L_x - F_x^2 Q^2 - L_x^2}{F_x^3 Q}) \quad (2)$$

$$f_{r1} = \frac{1}{2\pi\sqrt{C_r L_r}} \quad (3)$$

$$f_{r2} = \frac{1}{2\pi\sqrt{C_r(L_r+L_m)}} \quad (4)$$

$$F_x = \frac{f_s}{f_{r1}} \quad (5)$$

$$m = \frac{L_m + L_r}{L_r} \quad (6)$$

$$L_x = \frac{L_m}{L_r} \quad (7)$$

$$Q = \frac{\sqrt{L_r C_r}}{R_{ac}} \quad (8)$$

$$R_{ac} = \frac{8R_o}{\pi^2 N^2} \quad (9)$$

$$R_o = \frac{V_o^2}{P_o} \quad (10)$$

$$ZVS\ boundary, \delta = \theta - \frac{\varphi}{2} > 0 \quad (11)$$

Where,
N = secondary to the primary turns ratio of the transformer
$F_x$ = normalized switching frequency
$f_s$ = switching frequency
$f_{r1}$ = first resonant tank frequency
$f_{r2}$ = second resonant tank frequency
$L_m$ = magnetizing inductance of the transformer
$L_r$ = resonant tank inductance
$C_r$ = resonant tank capacitance
$R_o$ = output AC load resistance exhibited by the grid
$V_o$ = output RMS grid voltage
$P_o$ = output Power
$R_{ac}$ = load resistance reflected at the primary side of the transformer.

For buck or boost mode, N must be calculated properly. Buck mode is usually preferred due to guaranteed ZVS turn-on in primary side switches in the LLC converter and better control of gain with variation in frequency. PV inverters are usually designed to support a range of input voltages. For Grid-tied inverters, the output voltage is dictated by the grid voltage. To determine the maximum required gain, the minimum input voltage ($V_{in-min}$) with output grid tolerance percentage k (0-5%) depends on the grid voltage tolerance. With RMS grid voltage $V_o$, the required secondary to primary turn ratio N for one of the resonant converters for single-phase becomes,

$$N = \frac{(1+k)V_o}{\sqrt{6}V_{in-min}} \quad (12)$$

To design the $L_r$-$C_r$ pair, the first resonance frequency ($f_{r1}$) should be slightly below the minimum switching frequency ($f_{s-min}$) to ensure ZVS operation. After selecting the value of $f_{r1}$, multiple $L_r$-$C_r$ pairs can be selected for the same resonance frequency. Increasing the value of $C_r$ decreases the voltage stress on the resonant capacitor while allowing smaller $L_r$ and $L_m$ values. [22] suggests an expression for the minimum value of $C_r$ is given by,

$$C_{r-min} = \frac{NI_m}{4f_{s-min}(V_{cr-max}-\frac{V_o}{N})} \quad (13)$$

Where,
$V_{cr-max}$ = maximum voltage stress
$V_o$ = output rms voltage
$I_m$ = rated peak output current
$f_{s-min}$ = minimum switching frequency ($f_{s-min}$)

Smaller $L_m$ and $L_r$ increase the magnetizing and resonant currents, respectively, resulting in increased losses. Hence, it is a trade-off between the voltage stress, losses, and market availability of resonant capacitors. The design of the range of Q is now fixed once $L_r$ and $C_r$ pair is selected since it now only depends on the load resistance $R_o$. For Solar PV inverters, the amount of load is defined by the MPPT algorithm and is also in a range. Hence, the Q factor is in a range of values. As discussed above, $L_m$ should be large to decrease the transformer's conduction and circulating current loss. However, at low-load conditions, larger $L_m$ decreases the phase angle of the impedance and hence decreases the ZVS angle boundary as well, according to (11). Hence, switching losses are increased.

Moreover, larger $L_m$ increases the m-value, which decreases output voltage regulation with switching frequency modulation. Fig. 3. Shows Impedance angle variation with magnetizing inductance at different frequencies at different load conditions.

III. FILTER COMPONENTS DESIGN

Filter components of this system must be chosen carefully as well. Since the capacitor bank is connected to MV, the required capacitance at a given load current decreases. The value further decreases for transferring minimum power from PV to the grid. As the capacitor (Cf) and the inductor (Lf) are separated only by an unfolding inverter between them, it acts as an L-C filter. The calculation of the LC filter is given by,

$$C_f = \frac{0.1 I_{m-min}}{2\pi V_m f_g} \quad (14)$$

$$L_f = \frac{1}{4\pi^2 f_c^2 C_f} \quad (15)$$

Where,
$I_{m-min}$ = peak grid current at minimum power
$V_m$ = peak grid voltage
$f_g$ = grid frequency
$f_c$ = resonance frequency of the filter

The filter inductor (Lf) value should be chosen with the resonance of $L_f$ and $C_f$ in mind. The resonance frequency of the filter should not interfere and stay away from both the resonance frequencies $f_{r1}$ and $f_{r2}$. Cut-off Choosing $f_c \ll f_{r1}$ since above $f_{r1}$, there will be switching frequency components. There are two regions of operation: (1) $f_c > f_{r2}$ and (2) $f_c < f_{r2}$. In region 1, the operation at high load is improved. However, low-load operations have larger Q-factors, and their resonance tends towards $f_{r2}$. Therefore, the $f_c$ must be chosen such that it's far enough from $f_{r2}$ so that it doesn't interfere with the frequency components already excited by the converter resonance to improve low-load operations. The transient behavior and zero-crossings of the current waveforms improve in this region due to the higher cut-off frequency of the filter. However, THD is high at low load.

In region 2, when $f_c$ is far enough from $f_{r2}$, the operation of both Low and high-load operations can be improved performance, and the THD gets lowered. Since the filter's cut-off frequency is low and capacitor size is limited, the required inductor size, in this case, increases. This leads to an increase in the physical size of the system as well. To decide between these regions, trade-offs between THD, physical size, step-load behavior, and zero-crossing of the current waveform should be considered.

## IV. SYSTEM DESIGN METHODOLOGY

Based on the above discussion, a design procedure is proposed here, as shown in Fig 4. First, after getting the parameters, turns ratio N is calculated using (12), and the resonant components $C_r$ and $L_r$ are selected based on fs-min. The initial $C_r$ value is kept at $C_{r-min}$ as calculated in (13) to facilitate larger $L_m$ and $L_r$ values and decrease the losses. After the $L_m$ and Q-factor range selection, the converter is simulated in MATLAB Simulink, and the voltage across $C_r$ is observed. If it exceeds the maximum allowable voltage rating as per availability in the market, the $C_r$ is increased, and the resonant tank parameter selection is performed and simulated anew. If the voltage across $C_r$ is within the desirable range, The graphs M and θ vs. switching frequency for various Q factors within the viable range are plotted to check the output voltage regulation in the desired switching frequency range as shown in Fig 2 using (1) and (2). If the voltage regulation isn't achieved, especially for high loads, the $L_m$ value is decreased. Once the output voltage regulation across the desired switching frequencies is achieved, the losses are calculated from the simulated currents and voltages. The decrease in $L_m$ at low loads increases the ZVS range and is almost constant at high loads, as shown in Fig. 3. Hence, the losses at low loads are observed. If the losses are more than desired, the $L_m$ value is decreased. Otherwise, the filter components $L_f$ and $C_f$ are calculated. Depending on whether the designer needs to keep the filter inductance size compact, either region 1 or 2 can be chosen. Since region-2 produces a smoother sinusoidal current waveform due to higher inductance, this region is chosen if there is flexibility for size constraints of the filter inductor $L_f$. On the contrary, stricter size constraints would require region 2 operation. In both cases, the THD is simulated in MATLAB. If the THD is less than the maximum allowable requirement of $THD_{max}$, the design is complete. Otherwise, to lower the THD, the $L_f$ value is increased in region 2 (fc < fr2) to push the resonance frequency of the filter further away from $f_{r2}$. However, too much inductance could cause the filter resonance components to be close to grid frequency. Hence, the lower limit to be $Xf_g < f_c$ is applied here, where X is between 10 to 20. In region 1 ($f_{r2} > f_c >> f_{r1}$), the inductor value is decreased to get the resonance frequency of the filter away from $f_{r2}$. In this region, too little inductance can cause the filter's resonance to be near $f_{r1}$. To limit this, $Yf_c < f_{r1}$ can be applied here, where Y is between 2 to 5.

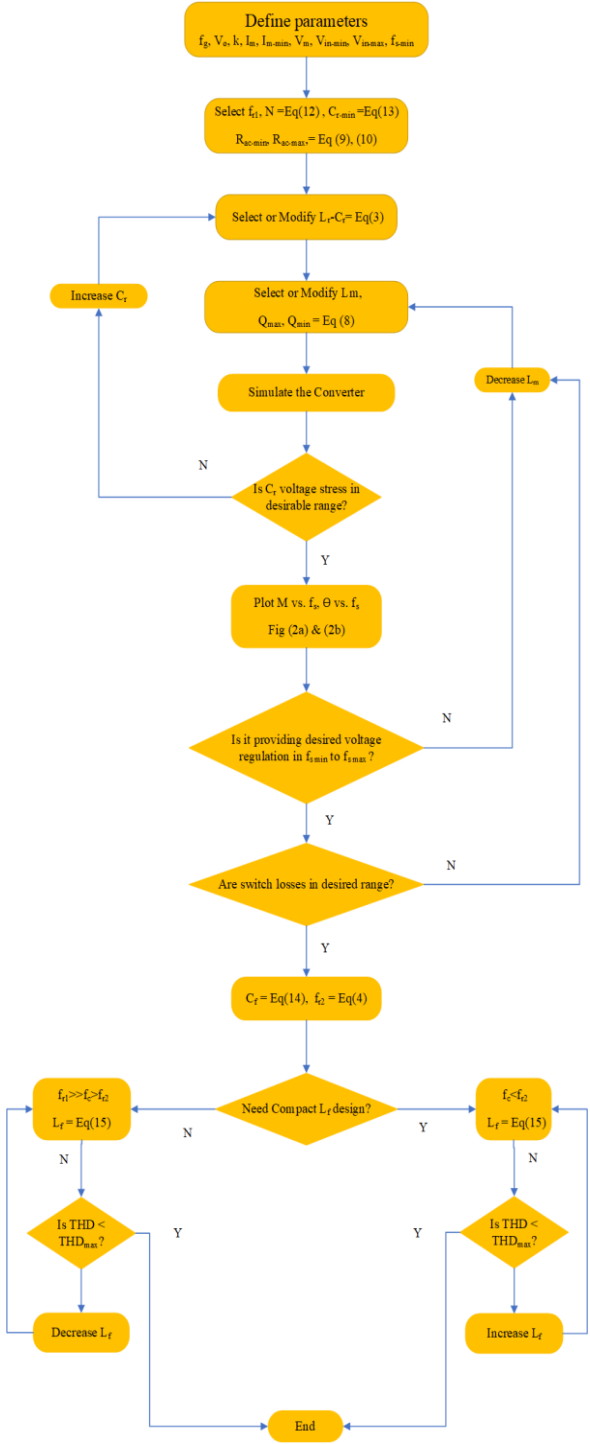

Fig. 4. Proposed Design Methodology.

TABLE I. CONVERTER PARAMETERS

| Parameters | Value |
| --- | --- |
| Rated Output Power (Single-phase) | 30kW - 1/3 MW |
| Grid Voltage and Frequency, $V_o$, $f_g$ | 13.8kV, 60Hz |
| Grid Voltage Tolerance, k | +/-5% |
| Input Voltage, $V_{dc}$ | 600 - 850V |
| Switching Frequency, $f_s$ | 45 - 70kHz |
| First Resonance Frequency, $f_{r1}$ | 44kHz |
| Maximum Resonance Capacitor Voltage, $V_{cr-max}$ | 850 V |
| Resonance Inductor. $L_r$ | 4µH |
| Resonance Capacitor, $C_r$ | 3.3µF |
| Magnetizing Inductor, $L_m$ | 100µH |
| Second Resonance Frequency, $f_{r2}$ | 10kHz |
| Secondary to Primary Turn Ratio, N | 10 |

## V. SIMULATION RESULTS

This methodology calculates the inverter parameters for maximum and minimum power points and selects the optimum solution. The parameters are designed in MATLAB and simulated through Simulink. A simple hybrid phase-shift and switching frequency modulation is shown in Fig.4, with Design Parameters shown below in Table I. The converter simulation has been performed as shown in Fig. 4 and presented with the Conditions (C1) 600V DC input for 0-0.0708s, (C2) 850V DC input for 0.0708-0.1625s, (C3) 700V DC input for 0.1625-0.254s (C4) 750V DC input for 0.254-0.3s with 60A peak Grid current output for 1/3 MW single phase system. C1 and C2 are cases with minimum and maximum inputs to the system, and C3 and C4 are in-between cases. During C1, the system's gain must peak during the peak grid output voltage. Hence, the phase-shift angle reaches 0, and the switching frequency nears the resonant frequency $f_{r1}$ at peak voltage. On the contrary, the voltage gain goes to zero at zero-crossings with phase-shift $\pi$, and the switching frequency is kept around the maximum possible frequency. The upper-frequency limit is decided by the primary side switch frequency as well as the transformer core. In this condition, the ZVS is achievable for a wide range of the sinusoid period due to low switching frequency and phase-shift angle. Hence, this is the best case in terms of switching losses. On the contrary, C2 demands the switching frequency and phase-shift angles to be higher to match the gain requirement. Increased switching frequency allows lower phase-shift angles for the same gain, and ZVS can be achieved over a wider range. However, compared to C1, C2 offers a narrower range of ZVS throughout a sinusoid and is the worst case in terms of switching losses. C3 and C4 are the in-between cases and offer a moderate range of ZVS. With this in mind, another set of conditions (C5, C9) Grid peak current demand Ipeak =50A for 0-0.0708s, (C6, C10) Ipeak = 10A for 0.0708-1625s, (C7, C11) Ipeak= 60A for 0.1625-0.254s and (C8, C12) Ipeak =30A for 0.254-0.3s is simulated with different loads for the worst case at 850V input with Parameter set in Table II for filter cut of frequency in region-1 ($f_{r1} >> f_c > f_{r2}$) and region-2 ($f_c < f_{r2}$) respectively.

Comparing the two filter designs, the system exhibits a THD of 10.35% for C6 and 4.53% for C7 for low-load conditions. The THD are 2.41% and 1.55% at high load conditions at C7 and C11, respectively. It is evident that C6 and C10 are the worst-case scenarios regarding THD, whereas Conditions C7 and C11 are the best. In region 2, THD improves overall due to increased filter inductance. The comparison also reveals filter resonance excitement at low loads. The transient response of the current waveform worsens as the current requires more time to stabilize once resonance frequencies are excited due to step changes shown in Fig. 6. The zero-crossings of the current waveform worsen in region-2 as well. Hence, if the designer needs to improve the transient response and zero-crossing, it is recommended to make use of region-1. On the contrary, if the need is to reduce THD, it is recommended to use region 2.

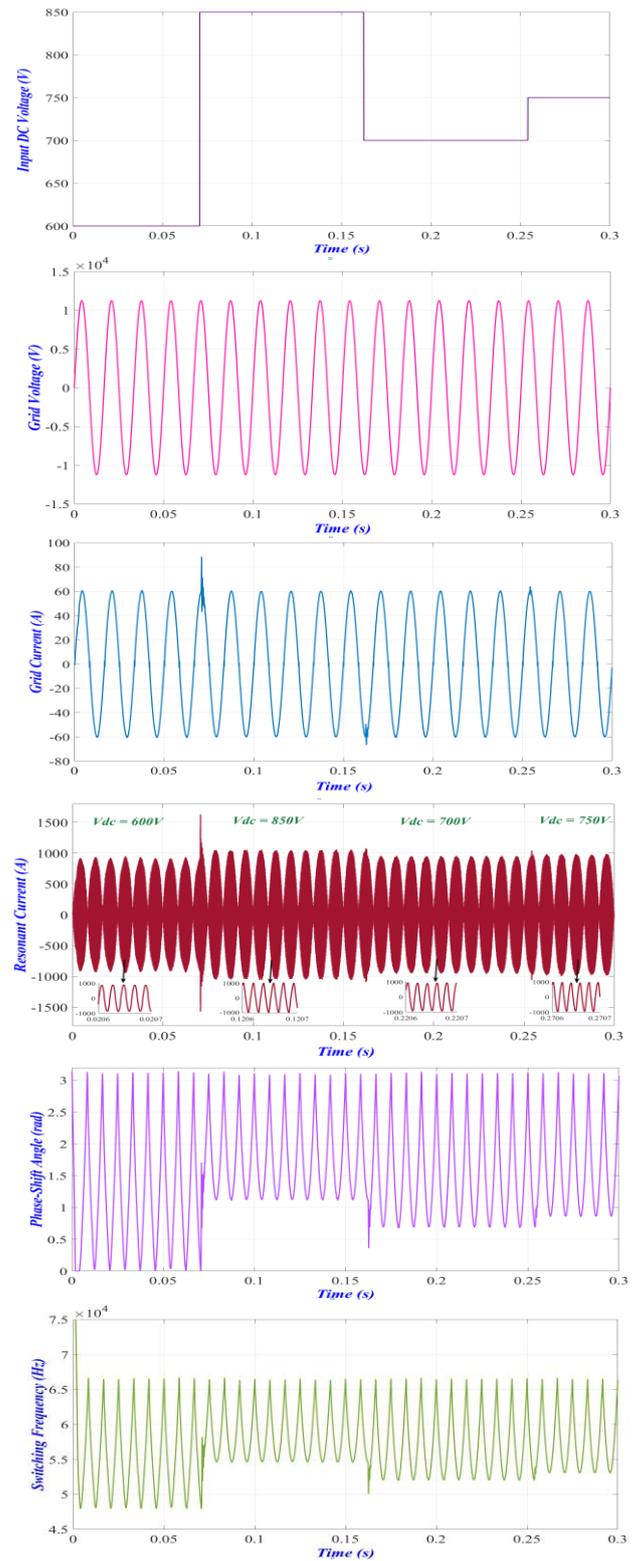

Fig. 5. System Waveforms at (C1) 600V DC input for 0-0.0708s, (C2) 850V DC input for 0.0708-0.1625s, (C3) 700V DC input for 0.1625-0.254s (C4) 750V DC input for 0.254-0.3s with 60A peak Grid current output for 1/3 MW single phase system.

TABLE II. FILTER COMPONENTS

| Filter Design Parameters Region-1 ($f_{r1}>>f_c>f_{r2}$) | | Filter Design Parameters Region-2 ($f_c<f_{r2}$) | |
|---|---|---|---|
| Cut-off Frequency, $f_c$ | 4kHz | Cut-off Frequency, $f_c$ | 13kHz |
| Filter Capacitor, $C_f$ | 0.1µF | Filter Capacitor, $C_f$ | 0.1µF |
| Filter Inductor, $L_f$ | 16mH | Filter Inductor, $L_f$ | 1.5mH |

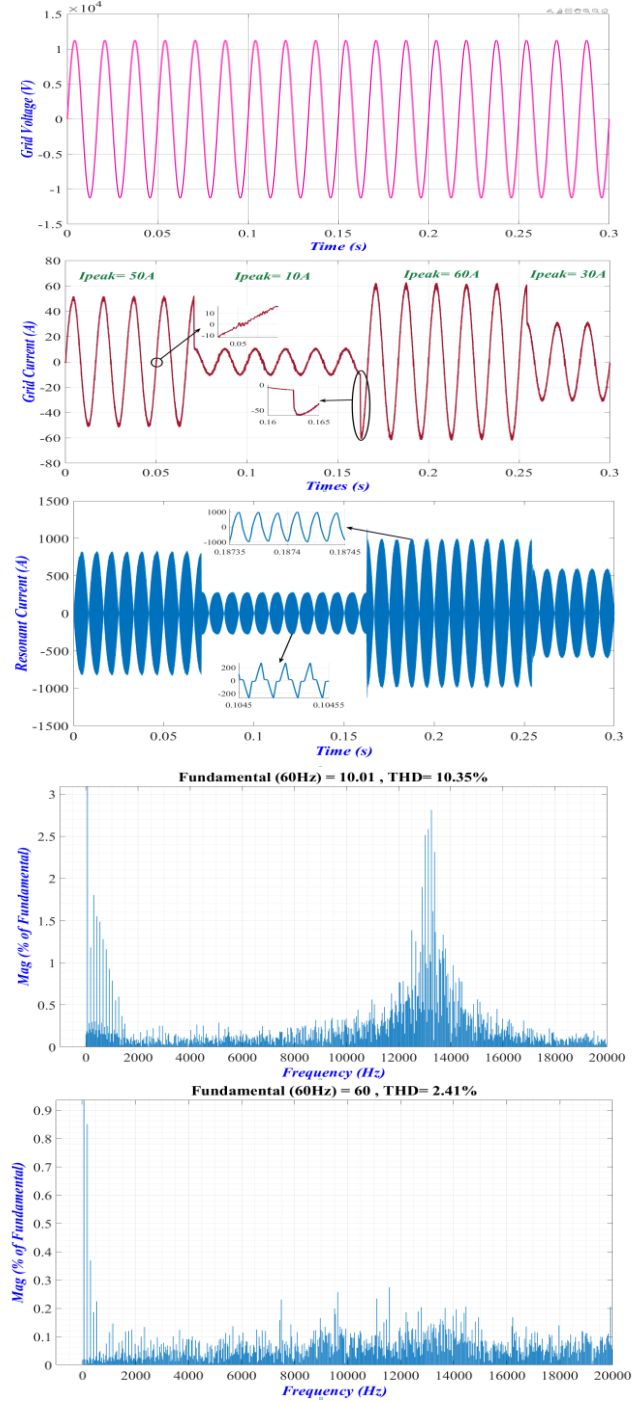

Fig. 6. System waveforms for (C5) Grid peak current demand Ipeak =50A for 0-0.0708s, (C6) Ipeak = 10A for 0.0708-1625s, (C7) Ipeak= 60A for 0.1625-0.254s and (C8) Ipeak =30A for 0.254-0.3s for filter design. Region-1 (fr1>>fc>fr2).

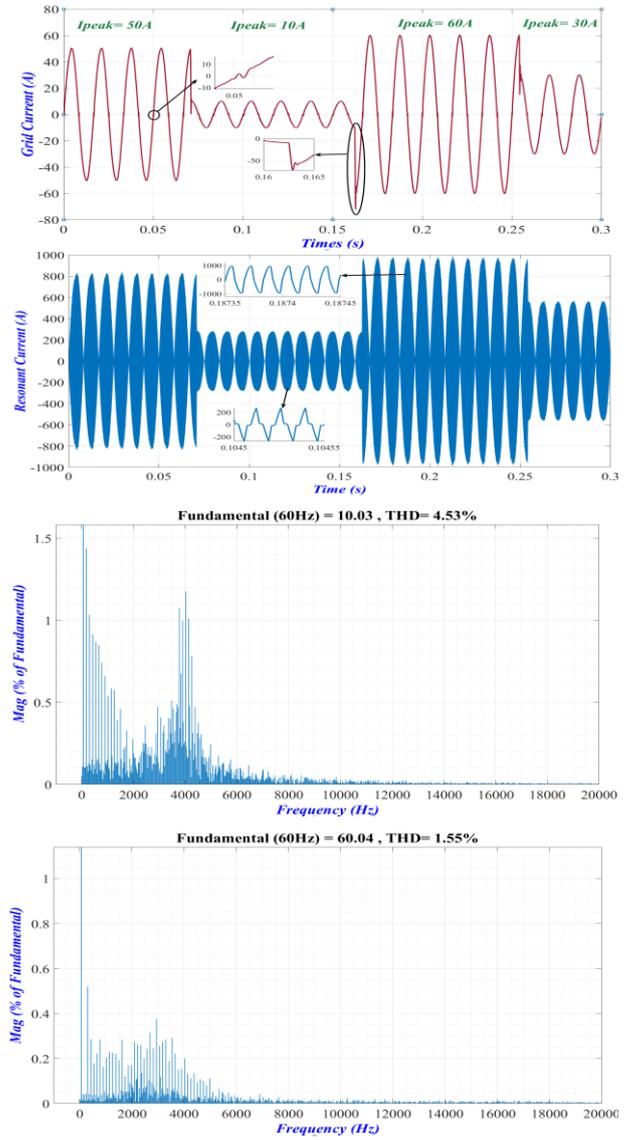

Fig. 7. System waveforms for (C9) Grid peak current demand Ipeak =50A for 0-0.0708s, (C10) Ipeak = 10A for 0.0708-1625s, (C11) Ipeak= 60A for 0.1625-0.254s and (C12) Ipeak =30A for 0.254-0.3s for filter design Region-2 (fc<fr2).

## CONCLUSION

This paper provides an MV LLC converter-based inverter design for PV systems. A unique design procedure for transformer turns ratio, leakage inductance, resonant capacitor, and magnetizing inductance has been proposed. The combination of phase-shift and switching frequency is discussed, and the impact on ZVS angle with increasing magnetizing inductance is considered for both high and low loads. In addition, a distinct design methodology for filter capacitors and inductors is also discussed. The design of the system is verified through simulations in MATLAB Simulink. The system is simulated at both the input voltage and output load variations and up to 1/3 MW single-phase power is injected successfully into the 13.8 kV single-phase grid pertaining to the 1 MW system requirement for a three-phase system. Two different filter designs are also simulated, and the filter performance is compared at low and high load conditions.